\documentclass[conference]{IEEEtran}
\IEEEoverridecommandlockouts
\usepackage[frozencache=true, finalizecache=false, cachedir=./minted-cache]{minted} 
\usepackage{cite}
\usepackage{amsmath,amssymb,amsfonts}
\usepackage{graphicx}
\usepackage{textcomp}
\usepackage{xcolor}
\usepackage[most]{tcolorbox}
\def\BibTeX{{\rm B\kern-.05em{\sc i\kern-.025em b}\kern-.08em
    T\kern-.1667em\lower.7ex\hbox{E}\kern-.125emX}}

\usepackage{tikz}
\usetikzlibrary{matrix, arrows}
\usepackage{amsmath,amssymb}
\usepackage{amsthm}
\usepackage{mathtools}
\usepackage{xspace}
\usepackage[noend]{algorithmic}
\usepackage[ruled,vlined]{algorithm2e}
\usepackage{url}
\usepackage{makeidx}
\usepackage{enumerate}
\usepackage{epstopdf}
\usepackage{booktabs}
\usepackage{color}
\usepackage[utf8]{inputenc}
\usepackage{thm-restate}
\usepackage{scalerel,stackengine}
\usepackage[shortlabels]{enumitem}
\usepackage{xr}
\usepackage{fancyvrb}
\usepackage{xcolor}
\usepackage{bold-extra}
\usepackage[width=474.18663pt]{caption}
\usepackage{subfigure}
\usepackage[most]{tcolorbox}
\usepackage{fvextra}
\usepackage[frozencache=true, finalizecache=false, cachedir=./minted-cache]{minted} 
\usepackage{float}
\usepackage{alltt}
\usepackage{soul}
\usepackage{fancyvrb}
\usepackage{multirow}
\usepackage[final]{hyperref}
\usepackage[bottom]{footmisc}
\usepackage{array}
\newcolumntype{P}[1]{>{\centering\arraybackslash}p{#1}}

\usepackage{listings}
\usepackage{tikz}
\usetikzlibrary{shapes,calc,positioning}

\global

\tcbset{
  aibox/.style={
    width=525.18663pt,
    top=10pt,
    colback=white,
    colframe=black,
    colbacktitle=black,
    enhanced,
    center,
    attach boxed title to top left={yshift=-0.1in,xshift=0.15in},
    boxed title style={boxrule=0pt,colframe=white,},
  }
}
\newtcolorbox{AIbox}[2][]{aibox,title=#2,#1}

\definecolor{aigold}{RGB}{244,210, 1} 
\definecolor{aigreen}{RGB}{210,244,211} 

\sethlcolor{aigreen}

\definecolor{aired}{RGB}{255,180,181}

\newtcbox{\mybox}[1][green]{on line,
arc=0pt,outer arc=0pt,colback=#1!10!white,colframe=#1!50!black,
boxsep=0pt,left=0pt,right=0pt,top=0pt,bottom=0pt,
boxrule=0pt,bottomrule=0pt,toprule=0pt}

\begin{document}

\title{Large Foundation Models for Power Systems
}

\author{    \IEEEauthorblockN{Chenghao Huang\IEEEauthorrefmark{1}\textsuperscript{\textsection}, Siyang Li\IEEEauthorrefmark{2}\textsuperscript{\textsection}, Ruohong Liu\IEEEauthorrefmark{2}\textsuperscript{\textsection}, Hao Wang\IEEEauthorrefmark{1}, and Yize Chen\IEEEauthorrefmark{2}}
    \IEEEauthorblockA{\IEEEauthorrefmark{1}Department of Data Science and Artificial Intelligence, Monash University, Australia
    \\\{chenghao.huang, hao.wang2\}@monash.edu}
    \IEEEauthorblockA{\IEEEauthorrefmark{2}Information Hub, Hong Kong University of Science and Technology (Guangzhou), China
    \\\{sli572, rliu519\}@connect.hkust-gz.edu.cn, yizechen@ust.hk}

}

\maketitle
\thispagestyle{plain}
\pagestyle{plain}
\begingroup\renewcommand\thefootnote{\textsection}
\footnotetext{Equal contribution.}
\endgroup

\begin{abstract}
Foundation models, such as Large Language Models (LLMs), can respond to a wide range of format-free queries without any task-specific data collection or model training, creating various research and application opportunities for the modeling and operation of large-scale power systems. In this paper, we outline how such large foundation model such as GPT-4 are developed, and discuss how they can be leveraged in challenging power and energy system tasks. We first investigate the potential of existing foundation models by validating their performance on four representative tasks across power system domains, including the optimal power flow (OPF), electric vehicle (EV) scheduling, knowledge retrieval for power engineering technical reports, and situation awareness. Our results indicate strong capabilities of such foundation models on boosting the efficiency and reliability of power system operational pipelines. We also provide suggestions and projections on future deployment of foundation models in power system applications.
\end{abstract}


\section{Introduction}
Recent advancements on large foundation models such as large language models (LLM) and large multi-modal models have greatly transformed the landscape of artificial intelligence and machine learning (ML)~\cite{bubeck2023sparks, radford2021learning}. The conversational variant of Generative Pre-trained Transformer 4 (GPT-4) developed by OpenAI, ChatGPT, has been widely perceived as a milestone toward large foundation models. While standard machine learning models are often-wise trained on single-source, domain-specific dataset, which is largely limited by the data collection and application scope. Large foundation models are based on the transformer neural network architectures~\cite{vaswani2017attention}, and are trained on unprecedented scale of compute and data. Once trained, they exhibit impressive, generalizable capabilities across a variety of domains and tasks including reasoning, abstraction, understanding and projection. Such models indeed revolutionize the development and implementation of machine learning algorithms and reveal many novel opportunities.

The equivalent- or even beyond-human-level abilities provided by foundation models motivate us to explore the emerging opportunities in boosting the performance of power system modeling and operations. \cite{powergpt} explores the pre-training of large transformer model for power system time-series forecasting and imputation, while in \cite{bonadia2023potential} the authors show early success on applying ChatGPT for generating test circuits in an interactive way. However, many abilities provided by foundation models are yet to be utilized for improving the reliability and efficiency of power networks. For instance, such large foundation models have been shown to be able to output understanding, reasoning, and using tools for a wide variety of domains to some degree~\cite{qin2023toolllm}. Such capabilities can potentially help or even change the operational procedures for large-scale power networks.

In this paper, we systematically develop application pipelines and examine the capabilities of large foundation models on a set of power system modeling and operation tasks. %We set up and collect working examples of large foundation models on typical applications ranging from power dispatch and EV scheduling, knowledge retrieval and question answering from long technical reports, situation awareness analysis with multimodal measurements, and energy timeseries forecasting. 
We find using off-the-shelf pre-trained foundation models such as GPT-4 and GPT-4 Vision (GPT-4V), it is indeed efficient to create multiple novel pathways in helping engineers, policymakers, and energy users for solving challenging problems in the power and energy domain. These problems often involve high-dimensional unstructured data, such as interacting with plain-language technical documents, system measurements or image data. In addition, by leveraging the power of large foundation models, it effectively removes the barriers for users to interact with the power systems. This differentiates from other machine learning implementation for power grid applications. For instance, previous research identified machine learning algorithms can successfully predict locational marginal price and power dispatch signals~\cite{chen2022learning, guha2019machine}. But this needs trained engineers to carefully craft labeled operational data and specific machine learning models. On the contrary, large foundation models can potentially ease the data collection and model training burden, while providing instant feedback for system operators and energy users~\cite{wu2022ai}.  To explore the capabilities provided  by large foundation models on power system tasks, we design specific prompts and interactive paradigms, and show foundation models can directly solve simplified optimal power flow (OPF) problems by only interacting with prompts without using optimization solvers; we also demonstrate LLMs can serve as an intermediary between end users and EV charging control algorithms, so that optimal scheduling models can be easily set up based on users' preferences; while pre-trained large foundation models can lack accurate knowledge in power and energy systems, we find that using retrieval-augmented generation~\cite{lewis2020retrieval} tackles such challenges by embedding large technical reports in specific domains.
%, such models can be readily available for reasoning and comprehension tasks for our domain-specific applications.

In particular, we focus on the unique characteristics of power system modeling and operations~\cite{kundur2007power}, as it exhibits varying complexity and domain knowledge, compared to other general natural language processing or vision-related tasks. The example applications shown in this paper can be extended to a variety of power system tasks, such as power grid modeling~\cite{kundur2007power}, market operations~\cite{kirschen2018fundamentals}, situation awareness~\cite{panteli2015situation}, energy management and power dispatch~\cite{roald2023power}, anomaly detection~\cite{ten2008vulnerability}, and forecasting tasks under various settings~\cite{shahidehpour2003market}. %Limitations of current foundation models in power system applications and possible future developments are also discussed. 
To facilitate the development and utilization of such powerful tools, we make our algorithms and detailed simulation testing cases, especially the multi-modal and additional time-series forecasting examples publicly available~\footnote{\url{https://github.com/chennnnnyize/LLM_PowerSystems}}.

\section{Methodology}
\begin{figure*}[htbp]
\centering
\includegraphics[width=0.65\textwidth]{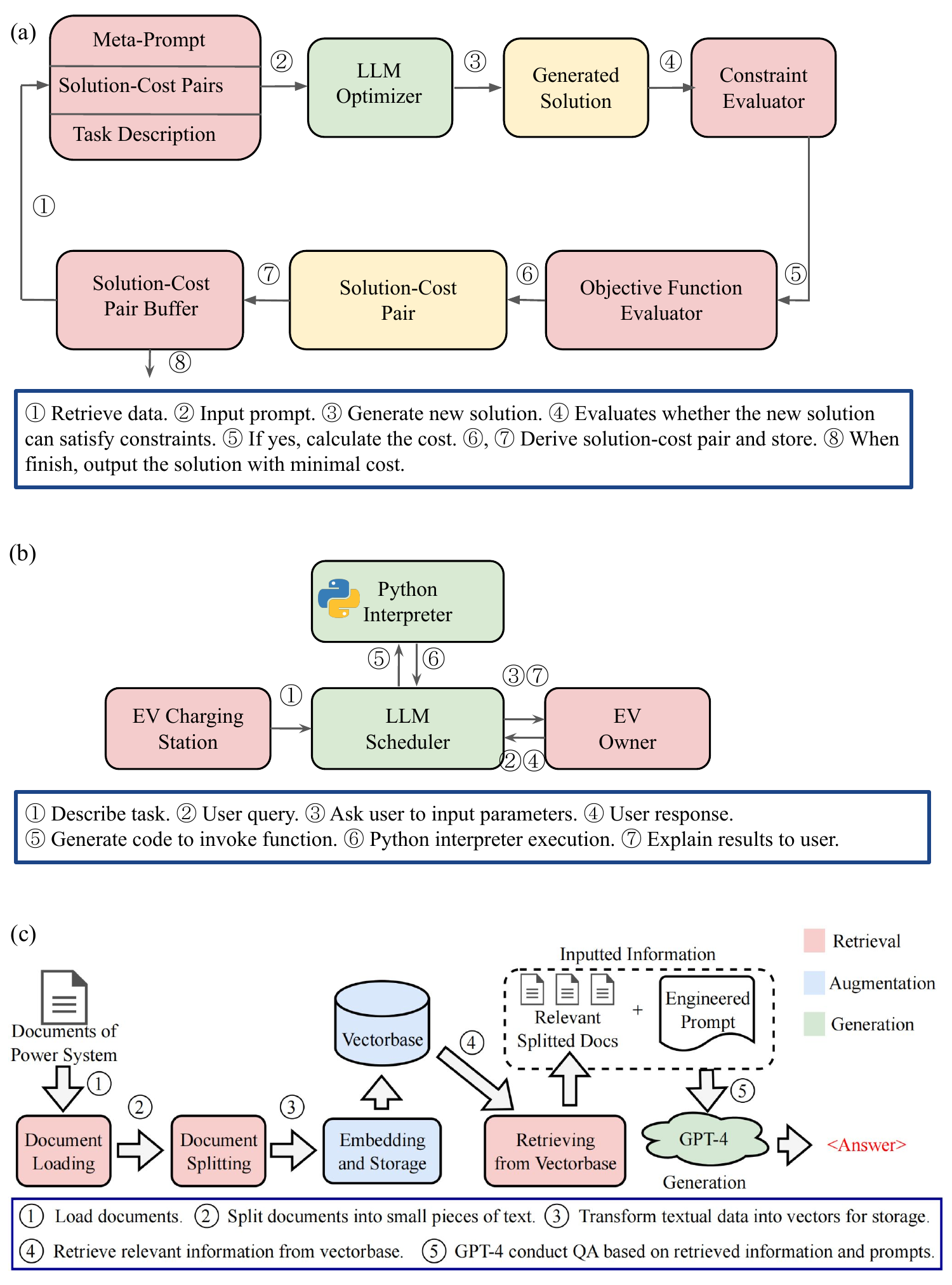}
\captionsetup{width=.99\linewidth}
\caption{Design of large foundation models on (a). Solving OPF (\texttt{LLM4OPF}); (b). Scheduling EV charging (\texttt{LLM4EV}); (c). Document summarization and knowledge retrieval. \vspace{-15pt}} 
\label{fig:method}
\end{figure*}

\subsection{LLM as Optimizer}
Optimization have been regarded as one of the foundational tools for solving various decision-making problems in power system, such as the OPF, EV charging scheduling, and building energy system management \cite{molzahn2017survey}. Although researchers have provided numerous solutions, there still exist challenges from both computation cost and algorithm performance perspectives. In particular, for a variety of users lacking domain knowledge in optimization or power systems, it is impractical for them to learn to model and solve such problems \cite{jin2023human}.

Optimization by PROmpting (OPRO) demonstrates the ability of LLMs to solve optimization problem by describing the problem with natural language, and then iteratively generates new solutions based on task description and previously generated solutions \cite{yang2023large}. We further develop a framework named LLM for the OPF (\texttt{LLM4OPF}) to solve constrained optimization problems as shown in Fig. \ref{fig:method}(a). In this framework, we generate prompt with  historical solution-cost pairs and task description. In each optimization step, several solution-cost pairs with minimal cost values are retrieved from solution-cost pair buffer, and LLM optimizer generates new solution based on the well-designed prompt. The new solution which satisfies physical constraints  will be stored in solution-cost pair buffer. The detailed prompt and response example is described in our online repository~\cite{online2023}. We additionally explore the ability of \texttt{LLM4OPF} to quickly adapt to new problems with different constraints, given new task description and solutions of previous problem.

Though OPRO can find good quality solutions in some optimization problems, the inefficiency in searching solutions space limits its application in optimization problems with larger scale. \cite{jin2023human} empirically shows the ability of LLMs to generate code, while occasionally triggering errors during execution. Hence, instead of relying on LLMs to generate the entire code, we introduce function templates for the optimization problem in the novel framework named LLM for EV (\texttt{LLM4EV}). LLMs interacts with end users, comprehends inquiries, writes code to invoke the functions, and explains the results of code execution to end users. Fig. \ref{fig:method}(b) illustrates our interactive process between users and LLMs. Complete interaction process is detailed in \cite{online2023}.

% \begin{figure}[htbp]
% \centerline{\includegraphics[width=0.5\textwidth]{Figures/llm4opf.png}}
% \caption{llm4opf} 
% \label{llm4opf}
% \end{figure}

% \begin{figure}[htbp]
% \centerline{\includegraphics[width=0.45\textwidth]{Figures/llm4ev.png}}
% \caption{llm4ev} 
% \label{llm4ev}
% \end{figure}

\subsection{LLM for Knowledge Retrieval}\label{section_LLM4QA}
In this subsection, we introduce our approach of adopting LLMs to achieve summarization and question answering (QA) on energy field-based long-textual documents, named \texttt{LLM4Doc}. The framework of our approach is depicted in Fig. \ref{fig:method}(c), where we adopt retrieval augmented generation (RAG) and prompt engineering (PE) as workhorses. RAG is a hybrid mechanism based on natural language generation and information retrieval. It is capable of enriching LLMs' knowledge by pulling in information from a corpus of useful data, much like fetching a book from a library \cite{lewis2020retrieval}. Integrated with RAG, LLMs can generate not only contextually accurate, but also information-rich outputs. \textbf{Retrieval} involves searching for relevant information from a large corpus of data, such as textual files and database. Only using retrieval can probably generate inaccurate or unrelated outputs.
\textbf{Augmentation} involves incorporating the retrieved information into LLM's internal representation of knowledge to significantly improve the output quality of LLMs.
\textbf{Generation} involves generating text from LLMs.
% Briefly, the principles of RAG can be broken down  as:
% \begin{enumerate}
%     \item \textbf{Retrieval} involves searching for relevant information from a large corpus of data, such as textual files and database. Only using retrieval can probably generate inaccurate or unrelated outputs.
%     \item \textbf{Augmentation} involves incorporating the retrieved information into LLM's internal representation of knowledge to significantly improve the output quality of LLMs.
%     \item \textbf{Generation} involves generating text from LLMs.
% \end{enumerate}

\begin{figure}[htbp]
\centering{\includegraphics[width=0.48\textwidth]{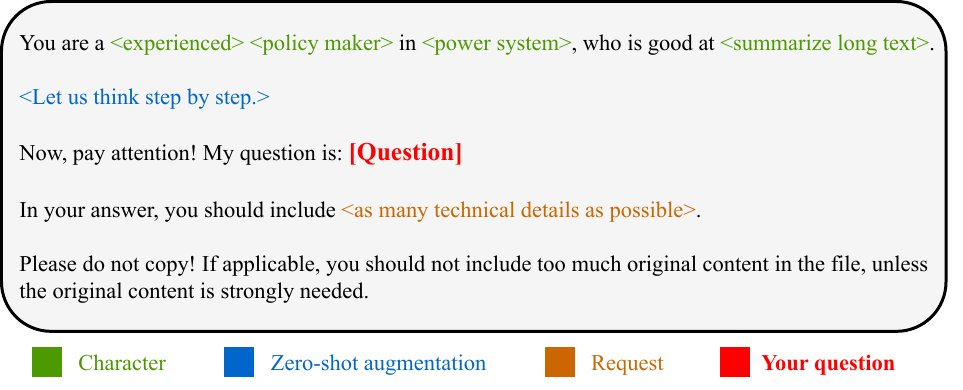}}
\captionsetup{width=.99\linewidth}
\caption{The  format of proposed  engineered prompts for processing technical documents in power systems. \vspace{-5pt}} 
\label{PE}
\end{figure}

We firstly process a document with long text into vector embeddings. Then, we use integrate these vectors into an external knowledge base. Based on this base, we ask GPT-4 to do summarization and multiple QAs on the processed document. On the other hand, prompt engineering is a crucial step in LLM development, which focuses on refining the language prompts to help LLMs understand human's requirements better, and produce fairly accurate, relevant, and information-rich outputs \cite{white2023prompt}. There are various purposes of prompt engineering. Specifically, in this paper, we design a format for our prompts, where humans can customize different parts to get diversified answers, detailed in Fig. \ref{PE}.

\subsection{Large Multimodal Models for Situation Awareness}
Multi-modal models are able to connect learned representations across diverse data types, e.g. language, vision, and etc, to gain a more comprehensive understanding and generate appropriate responses. Current state-of-the-art multi-modal models are built upon LLMs with incremental pre-training with image-text pairs \cite{tsimpoukelli2021multimodal}, so the resulting models can achieve a variety of tasks with multiple types of model input and output data. Such capabilities can provide valuable tools for power system applications, as system measurements and logging data are naturally large scale and multi-modal. 

In particular, we look into the situation awareness problem which can be tackled by large multi-modal models \cite{panteli2015situation}. To guarantee the operation safety, system operators need to make timely and efficient measurements processing, operating states comprehension, and anomaly detection. %An efficient way is to analyze real-time snapshots perceived by unmanned aerial vehicles (UAVs) using visual recognition methods. For instance, YOLOv2 is exploited in \cite{sadykova2019yolo} to identify the insulator surface status. 
For instance, imaging data have been used in \cite{sadykova2019yolo} to identify the insulator surface status. However, there are \textit{limited sample availability of target situations and limited diversity of operating situations}, %There exits a data imbalance of different situations for a studied object, 
such as the sub-station usually stays in normal stage with scare fire disaster occurrence, which render it hard to foster a dedicated fire detector. %2) \textit{situation diversity}. There could be a broad variety of intriguing conditions to recognize, which is really an obstacle to learn a generic monitor to account for all kinds of operating conditions. 3) \textit{background complexity}. The complex open environment behind the target object also raises disturbance for accurate detection. 
Hence, a zero/few-shot situation monitor with reliable and aligned vision-text representations is highly favored. GPT-4V is a competent candidate \cite{yang2023dawn}, which owns potent image-text retrieval ability to produce accurate scene descriptions for measured images, as well as outstanding zero/few-shot generalization inference and scenarios transferability to discern rare situations. In this work, we propose a framework named \texttt{LLM4SA} by adopting in-context learning plus prompt engineering \cite{cao2023towards} which circumvents cumbersome post-hoc fine-tuning. In this regard, proper prompt design plays a key role which should provide multi-modal QA exemplars of high correspondence and concrete detection task or situation category information \cite{min2022rethinking}.

There are two orthogonal ways to achieve desirable situation understanding. One way is instruction tuning coupled with prompt learning \cite{wei2021finetuned}, which needs to handcraft domain-specific input-label pairs and use them to tailor an ad-hoc prompt embedding and refine original parameters for LLMs. But this method is hard to execute due to formidable data collection and computational efforts for large parameter fine-tuning. The other way is in-context learning plus prompt engineering \cite{cao2023towards} which circumvents cumbersome post-hoc fine-tuning. In this regard, proper prompt design plays a key role which should provide multi-modal query-answer exemplars of high correspondence and concrete detection task or situation category information \cite{min2022rethinking}. In this work, we majorly focus on the in-context learning capabilities to attain the zero/few-shot situation awareness by curating detailed in-context prompts.

\section{Applications in Power Systems}

\subsection{Optimal Power Flow}
In this paper, we consider the OPF problem with simplified constraints on nodal power balance and power flow equations. The detailed formulation can be found in \cite{online2023}. In this problem, we minimize the quadratic power generation costs $C_i\left(P_i\right)=a_iP_i^2+b_i P_i+c_i, i \in \Omega$, where $P_i$ denotes power output of generator $i \in \Omega$. $a_i$, $b_i$ and $c_i$ denote cost coefficients. All generators have power generation limits.

We evaluate \texttt{LLM4OPF} on a 5-unit testing case \cite{soroudi2017power}, and compare its performance with Gurobi solver \cite{gurobi2021gurobi} on two tasks. In the first task, we set the sum of nodal load $L_e = 400$ and instruct GPT-4 to perform $n=300$ optimization steps. Subsequently in the second task, we set $L_e = 405$ and include the solution-cost pairs obtained from the first experiment as part of the prompt. We then tasked GPT-4 API to perform 50 optimization steps to evaluate the the ability of \texttt{LLM4OPF} to adapt to new tasks with different constraints.

\begin{figure}[htbp]
\centering{\includegraphics[width=0.5\textwidth]{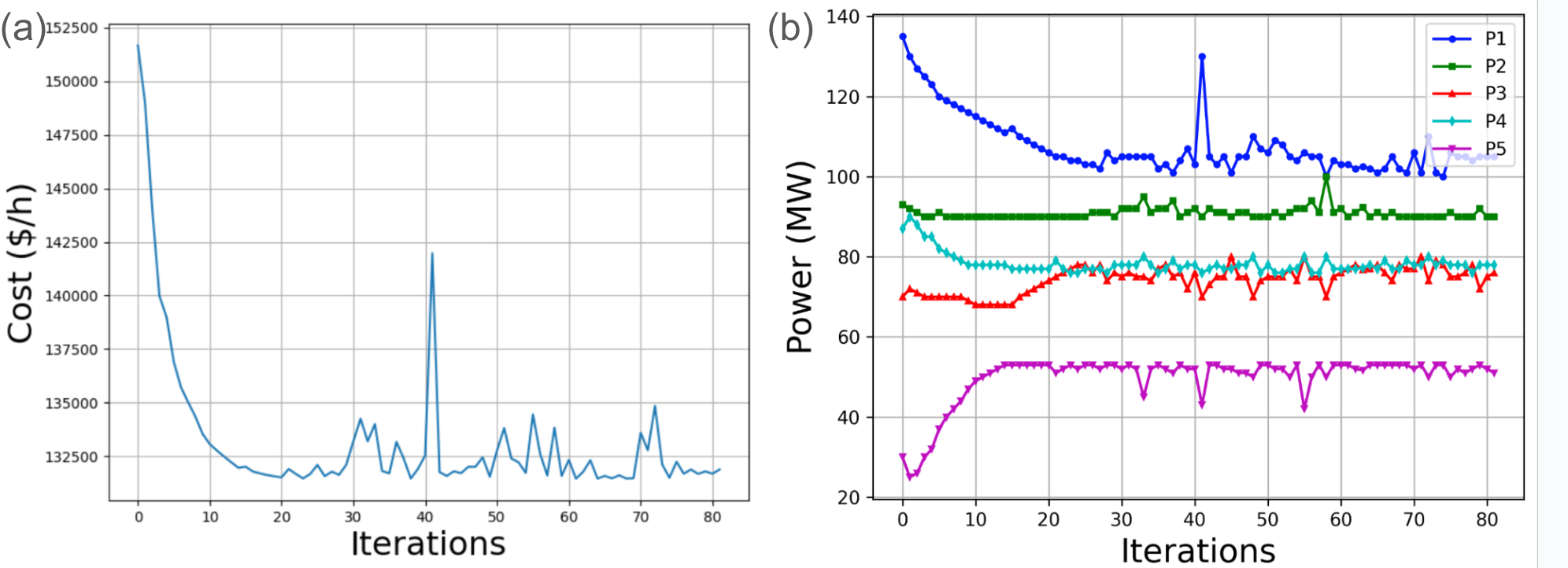}}
\captionsetup{width=.99\linewidth}
\caption{OPF testing case (a). total costs; (b). generation values.}
\label{opfresults}
\end{figure}

The experimental results depicted in Fig. \ref{opfresults}(a) and Fig. \ref{opfresults}(b) demonstrate that the \texttt{LLM4OPF} can iteratively optimize to minimize the cost. Note that we only store and plot the solutions that satisfy constraints. As shown in Table. \ref{opftable}, the cost and generator thermal output values obtained by \texttt{LLM4OPF} closely match those achieved by the Gurobi optimizer. Furthermore, it is evident that \texttt{LLM4OPF} can achieve the solution for the new task with fewer optimization steps when provided with the solution to relevant problems. These results show the strong capabilities of LLMs in solving optimization tasks.

% \begin{figure}[htbp]
% \centering
% \subfigure[]{\label{cost}\includegraphics[width=0.24\textwidth]{Figures/cost.png}}
% \subfigure[]
% {\label{power}\includegraphics[width=0.24\textwidth]{Figures/power.png}}
% \caption{5-unit OPF testing case experiment results.}
% \label{opfresults}
% \vspace{-8pt}
% \end{figure}

% Please add the following required packages to your document preamble:
% \usepackage{multirow}
\begin{table}[htbp]
\caption{Comparison of Experiment Results on EV Charging.}\label{opftable}
\resizebox{1\columnwidth}{!}{%
\begin{tabular}{ccccccccc}
\hline
Task                                                                         & Optimizer & Cost       & $P_1$      & $P_2$     & $P_3$     & $P_4$     & $P_5$     & n   \\ \hline
\multirow{2}{*}{OPF}                                                         & Gurobi    & 131455.000 & 102.844 & 90.000 & 76.730 & 77.425 & 53.000 & -   \\
                                                                             & GPT-4     & 131456.090 & 103.000 & 90.000 & 77.000 & 77.000 & 53.000 & 300 \\ \hline
\multirow{2}{*}{\begin{tabular}[c]{@{}c@{}}OPF\\$L_e = 405$\end{tabular}} & Gurobi    & 134670.416 & 104.850 & 90.000 & 78.216 & 78.934 & 53.000 & -   \\
                                                                             & GPT-4     & 134691.790 & 106.000 & 90.000 & 79.000 & 77.000 & 53.000 & 50  \\ \hline
\end{tabular}}
\end{table}

\begin{figure}
\begin{tcolorbox}[top=10pt, colback=white, colframe=black, colbacktitle=black, center, enhanced, breakable,
attach boxed title to top left={yshift=-0.1in,xshift=0.15in}, boxed title style={boxrule=0pt,colframe=white,}, title=LLM4EV]
{\bf System:} {\small You are an AI assistant specialized in solving EV charging problems. You have been provided with a predefined function called solve\_EV() that is capable of addressing various EV charging problems:
\begin{lstlisting}
def Solve_EV(num_of_vehicles, timesteps, initial_states, max_power, terminal_states, dept_time, power_capacity, plot_fig):
...
\end{lstlisting}
}

{\bf User:}\sethlcolor{green!20}\hl{Can you help me to schedule the charging of electric vehicles?}

{\bf Assistant:}{\small \sethlcolor{blue!20}\hl{Of course, I'd be happy to help. To provide the best result, I need some specific details from you. Could you please tell me the following:...}

What are the initial states (charge levels) of your vehicles? Please list for each vehicle if more than one...}

{\bf User:} {\small\sethlcolor{green!20}\hl{
3. They all start from zero, ...}}

{\bf Assistant:} {\small \sethlcolor{blue!20}\hl{Thank you for the information. Based on your input, we can utilize our EV charging solver as follows:}
\begin{lstlisting}
# starting from zero for all vehicles
initial_states = [0, 0, 0, 0, 0]
...
\end{lstlisting}}
\end{tcolorbox}
\caption{\texttt{LLM4EV} experiment results.\vspace{-15pt}}
\label{LLM4EV results}
\end{figure}

\subsection{Electric Vehicle Charging}
We consider a charging station which can provide charging service simultaneously for a set of EV $j\in\mathcal{V}$ \cite{chen2022state}. The electric vehicle charging optimization problem is formulated to optimize charging power $u_j(t)$ for each EV, subject to the EV charging dynamics $x_j(t)=x_j(t-1)+\delta u_j(t)$. $\delta$ and $x_j(t)$ denote the charging efficiency and EV state-of-charge at time $t$ respectively. We conduct experiment using a test case consisting of 5 charging sessions and 20 time steps to evaluate the performance of \texttt{LLM4EV} framework. Through this test, we primarily assess the performance of our proposed method in the following aspects:
\begin{itemize}
    \item Given task description and code example, can LLMs utilize natural language to ask user for required parameters to solve EV charging optimization problem?
    \item Can LLMs comprehend EV charging optimization problems described in natural language, and generate correct 
    code to call the provided functions?
    \item Can LLMs accurately interpret the output of the code and explain the results to the user?
\end{itemize}

From Fig.\ref{LLM4EV results}, we discover that GPT-4 can intelligently generate precise questions to users based on the provided function and task description. Remarkably, the 8 questions raised perfectly correspond to the 8 arguments of the function solve\_EV. Furthermore, even when users provide answers in natural language, GPT-4 demonstrates a strong ability to comprehend and generate corresponding code. For instance, when presented with the question ``what are the initial states (charge levels) of your vehicles? Please list for each vehicle if more than one", the user's answer is ``they all start from zero". Subsequently, GPT-4 generates accurate code 'initial\_states = [0, 0, 0, 0, 0]'. See \cite{online2023} for detailed experiment results.

% satellite image-based wildfire detection
% Fire disaster detection for real-world sub-station operating scene

% \cite{wildfiredataset}
\subsection{Situation Awareness via Multimodal Capabilities}
We utilize multimodal capabilities of GPT-4V to achieve situation awareness based on scenario images. Situation awareness usually involves classification tasks such as anomaly detection. In this part, we demonstrate the results of satellite image-based wildfire detection \cite{wildfiredataset}, a binary classification task. We design 4 approaches: (1). One image query and directly asking if wildfire exists in the image;
(2). One image query and applying prompt engineering;
(3). Multiple image examples along with ground true labels as the engineered prompts;
(4). Based on (3), adding language explanations associated with true labels.
% \begin{enumerate}
%     \item One image query and directly asking if wildfire exists in the image;
%     \item One image query and applying prompt engineering;
%     \item Multiple image examples along with ground true labels  as the engineered prompts;
%     \item Based on 3), adding language explanations for associated with true labels;
% \end{enumerate}

\begin{figure}
\begin{tcolorbox}[top=10pt, colback=white, colframe=black, colbacktitle=black, center, enhanced, breakable,
attach boxed title to top left={yshift=-0.1in,xshift=0.15in}, boxed title style={boxrule=0pt,colframe=white,}, title=LLM4SA]
{\bf Image inputs:}\\
\includegraphics[width=0.9\textwidth]{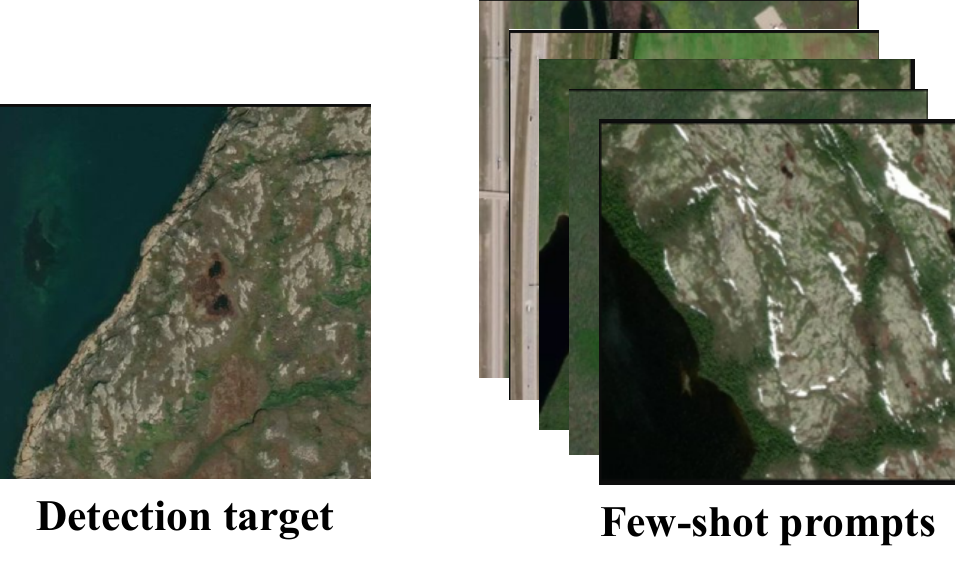}

{\small {\bf System:}} 
{\small This dataset contains satellite images about wildfire in Canada, using Longitude and Latitude coordinates for each wildfire spot ($>$ 0.01 acres burned) found. Areas after wildfire may demonstrate different color in satellite images.} \\
{\small You are a professor of forestry, and good at observing satellite images. I will give you several examples of satellite images with ``yes'' or ``no'' to specify if wildfire happened. The truths of the first 5 images are ``yes'', ``yes'', ``yes'', ``no'', ``no''.} \\
{\small {\bf Question:} \sethlcolor{green!20}\hl{Now, let's think step by step, and tell me, had wildfire happened in the last picture?}} \\
{\small {\bf Answer:} \sethlcolor{blue!20}\hl{...Considering this observation, it is likely that the last picture depicts an area where wildfires have occurred.}}\\

\end{tcolorbox}
\caption{An example of \texttt{LLM4SA} inference.}
\label{fig:LLM4SA}
\end{figure}

To get average accuracy of each approach, we take multiple-round experiments, where 5 positive images and 5 negative images are involved in each round. An example of Approach (3) is demonstrated in Fig. \ref{fig:LLM4SA} and see  \cite{online2023} for complete results. Among all approaches, the Approach 4) achieves the highest accuracy 91\%, proving that GPT-4V is capable of utilizing multimodal information to improve performance, while the Approach (3) achieves 86\%, also indicating the effectiveness of few-shot prompting. However, it is worth mentioning  GPT-4V's performance on specific image classification tasks is still inferior to standard ML models such as convolutional neural network with 94\% accuracy, and random forest with 96\% accuracy. Even so, GPT-4V holds remarkable potential in saving the training time and data collection processes, which are needed for typical ML methods. Furthermore, GPT-4V have more flexibility on accepting and generating multimodal outputs, such as generating explanations of current states.

% \begin{figure*}[htbp]
% \centering
%     \begin{AIbox}{LLM4SA}
%     \parbox[t]{0.3\linewidth}{
%     {\small {\bf Satellite images:}}\\
%     }
%     \hspace{0.2in}
%     \parbox[t]{0.3\linewidth}{
%     {\small {\bf System:}}\\
%     {\small This dataset contains satellite images about wildfire in Canada. This dataset uses Longitude and Latitude coordinates for each wildfire spot (> 0.01 acres burned) found on the dataset.} \\
%     {\small Tips: Areas with wildfire may demonstrate different color in satellite images.} \\
%     {\small Now I will give you several examples of satellite images with 'yes' or 'no' to specify if wildfire exists. The truths of the first 5 images are 'yes', 'yes', 'yes', 'no', 'no'.} \\
%     {\small You are a professor of forestry, and good at observing satellite images.} \\
%     {\small {\bf Question:} \sethlcolor{green!20}\hl{Now, let's think step by step, and tell me, did wildfire happen in the last picture?}} \\
%     \tcbline
%     {\small\bf Answer:} \scriptsize \begin{alltt}
%     \sethlcolor{blue!20}\hl{...Considering this observation, it is likely that the last picture depicts an area where wildfires have occurred.}
%     \end{alltt}}\\
%     \end{AIbox}
%     \caption{LLM4SA experiment results.}
%     \label{fig:LLM4SA}
%     \end{figure*}

\subsection{Energy Context-based Document Comprehension}

\begin{figure*}[htbp]
\centering
    \begin{AIbox}{LLM4Doc}
    \parbox[t]{0.5\linewidth}{
    {\small {\bf Directly ask questions.}}\\
    {\small {\bf Question:} \sethlcolor{green!20}\hl{What is Synchronization?}}\\
    {\small\bf Answer:} \scriptsize 
    \sethlcolor{blue!20}\hl{The synchronization requires generators to maintain synchronism where possible and continue to inject current to support system stability.}
    }
    \hspace{0.2in}
    \parbox[t]{0.5\linewidth}{
    {\small {\bf Directly ask questions.}}\\
    {\small {\bf Question:} \sethlcolor{green!20}\hl{What is phase-lock loop Synchronization?}}\\
    {\small\bf Answer:} \scriptsize 
    \sethlcolor{blue!20}\hl{I don't know.}
}
    \tcbline
    \parbox[t]{0.5\linewidth}{
    {\small {\bf Ask based on RAG.}}\\
    {\small {\bf Question:} \sethlcolor{green!20}\hl{What is Synchronization?}}\\
    {\small\bf Answer:} \scriptsize 
    %\begin{alltt}
    \sethlcolor{blue!20}\hl{... \textbf{synchronization} in this document refers to the synchronization status between the inverter and the grid during Bulk-Power System (BPS) events. \textbf{Specifically}, when a BPS event (such as a fault) causes instantaneous synchronization loss, ... \textbf{The purpose of synchronization} is to ensure that the inverter can continue to provide power and resynchronize with the grid...}
    %\end{alltt}
    }
    \hspace{0.2in}
    \parbox[t]{0.48\linewidth}{
    {\small {\bf Ask based on RAG.}}\\
    {\small {\bf Question:} \sethlcolor{green!20}\hl{What is phase-lock loop Synchronization?}}\\
    {\small\bf Answer:} \scriptsize 
    %\begin{alltt}
    \sethlcolor{blue!20}\hl{... \textbf{Specifically, PLL synchronization refers to} the inverter continuing to inject grid current during system disturbances, locking the phase-locked loop to the last synchronization point, and continuing to inject current into the grid at its calculated phase until the phase-locked loop can resynchronize. ... \textbf{maintaining PLL synchronization is important for} protecting and controlling ...}
    %\end{alltt}
    }
    \end{AIbox}
    \caption{Evaluation of two technical question answerings based on FERC Order No. 901 with and without RAG. \vspace{-5pt}}
    \label{fig:LLM4Doc:TQ12}
    \end{figure*}

\begin{table*}[htbp]
\scriptsize
\centering
\caption{Summary of use cases, limitations and possible future developments }
\begin{tabular}
{P{2cm}|m{3cm}m{4.5cm} m{7cm}}
\hline
\textbf{Tasks}       & \textbf{Limitations} & \textbf{Description} & \textbf{Potential Solutions} \\ \hline
OPF & Expensive computation  & Takes multiple iterations to converge & Improved initialization and few-shot prompts design  \\ 
 & Sample-specific optimization  & Only work on sample-by-sample load    &   Few-shot learning and reasoning across batch of system conditions \\ 
  \hline
 EV    &Code template dependent   &When changing problem formulation, new functions are required   &Expand function library for various problem formulations   \\ 
 \hline
Knowledge & Answer inconsistency & Generate different but useful answer sets& Ensemble learning and improved alignment \\ 
Retrieval &
Lack of specific knowledge & Limited to given documents & Fine-tuning on domain specific or private datasets    \\
 \hline
 Situation Awareness  &   Unstructured and multi-source input data, e.g., meter readings, image snapshots   &  Measurements come from heterogeneous devices and data sources, making it hard to align for foundation models  &          Different modalities, such as PMU measurements and images, provide different ``views" that complement each other, and need to be integrated to facilitate a holistic understanding                    \\ \hline
\end{tabular}\label{table:limitations} 
\end{table*}

In this experiment, we construct a pipeline shown in Sec~\ref{section_LLM4QA} for profiling and retrieving knowledge from power system documents, and apply it on the 177-page Federal Energy Regulatory Commission (FERC) Order No. 901~\cite{FERC}. This FERC order develops reliability standards addressing reliability gaps with inverter-based resources involved. %containing over 40000 words, 260000 characters.
To explore GPT-4's capability of different QA tasks, we classify questions into 2 types.
\textbf{General questions} ask about the overall content of the document, such as ``Please summarize this file.", ``What is the structure of this technical report?". While \textbf{Technical questions} focus more on the technical details mentioned in the document, such as ``What is the phase-lock loop Synchronization in this file?",  ``Please tell me, in Reliability Standards in this file, which standard is the most useful one?" Complete evaluation questions list is in \cite{online2023}. To study the professional capability brought by RAG, we compare the answers on two \textbf{technical questions} as shown in Fig. \ref{fig:LLM4Doc:TQ12}. GPT-4 with RAG is capable of handling deeper comprehension of large energy context-based documents. We can see that without RAG, GPT-4 but cannot understand relatively complex technical phrases. After using RAG, GPT-4 can not only interpret technical words in details, but also explain their importance or purpose. This helps understanding technical reports as well as embedding user-specific power system knowledge into foundation
 models.

% \vspace{-5pt}
\section{Limitations, Summary and Future Work}
In this paper, we explore a set of promising applications, promises and challenges of developing foundation models for power systems. The potential of such models are validated by a range of comprehension, modeling, and operation tasks. We show the noteworthy performance provided by LLMs and multi-modal models without fine-tuning or further training. We conclude that such strong capabilities of foundation models can provide system operators, policymakers, and end users efficient tools. 

While we note that large foundation models are still at the initial stage for industrial deployments, since there still exists  limited understanding to the model's behaviors, and needs thorough evaluation and testing on real-world scenarios. We also notice a set of limitations on using the current foundation models, such as model validity, efficiency and applicable scenarios, as detailed in Table \ref{table:limitations}.

%\textcolor{red}{Need to discuss limitations of the techniques, better shape as a complementary or helper for current task.}

%\textcolor{red}{require post-training adjustments to enforce business rules, rectify undesired behavior, and align with user values.}

\bibliographystyle{IEEEtran}
\bibliography{bib}

\appendix
\label{appendix}

\section{Experiment Setup}
\subsection{Optimal Power Flow}
\label{appendix_OPF}
We consider a simple economic dispatch problem of thermal units, which transform the fuel-based source of energy into
electricity \cite{soroudi2017power}. The thermal unit ecomomic dispatch problem is formulated as follows:
\begin{subequations}
\begin{align}
\min \quad & \sum_{i \in \Omega} C_i\left(P_i\right) \\
s.t.\quad  & C_i\left(P_i\right)=a_i\left(P_i\right)^2+b_i P_i+c_i, i \in \Omega \\
& P_i^{\text{min}} \leq P_i \leq P_i^{\text {max}}, i \in \Omega \\
& \sum_{i \in \Omega} P_i \geq L_e
\end{align}
\end{subequations}
In this problem, we optimize over power output $P_i$ of generator $i \in \Omega$ to minimize generation costs calculated in (1b). $a_i$, $b_i$ and $c_i$ denotes cost coefficients. (1c) and (1d) are operating limits and total generation constraints, respectively. $P_i^{ \text { min }}$ and $P_i^{ \text { max }}$ represents maximum/minimum power outputs of thermal unit $i$. $L_e$ is demand.

\begin{tcolorbox}[top=10pt, colback=white, colframe=black, colbacktitle=black, center, enhanced, breakable,
attach boxed title to top left={yshift=-0.1in,xshift=0.15in}, boxed title style={boxrule=0pt,colframe=white,}, title=LLM4OPF]
{\bf System:} You need assistance in solving an optimization problem. This problem involves 5 optimization variables, namely p1, p2, p3, p4, and p5. These variables are subject to constraints defined by their minimum and maximum values: $p_{min}$=[28, 90, 68, 76, 19] and $p_{max}$=[206, 284, 189, 266, 53]. Additionally, the sum of p1, p2, p3, p4, and p5 must be greater than or equal to 400. \\
Your objective is to provide values for p1, p2, p3, p4, and p5 that satisfy the constraints and minimize the optimization objective. \\
Below are some previous solution and their objective value pairs. The pairs are arranged in descending order based on their function values, where lower values are better.\\
\begin{lstlisting}
Solution 1: p1, p2, p3, p4, p5 = 120.0, 90.0, 70.0, 85.0, 40.0\\
Objective value 1: 141843.15\\
Solution 2: p1, p2, p3, p4, p5 = 110.0, 95.0, 75.0, 85.0, 35.0\\
Objective value 2: 137350.95\\
...

\end{lstlisting}
Give me a new (p1, p2, p3, p4, p5) pair that is different from all pairs above, and has a function value lower than
any of the above. Do not give me any explanation, the form of response must strictly follow the example: p1, p2, p3, p4, p5 = 123.11, 80.2, 99.67, 101.52, 37

\end{tcolorbox}

\subsection{Electric Vehicle Charging}
\label{appendix_EV}
The full EV charging scheduling problem is formulated as follows:
\begin{subequations}
\begin{align}
\min _{\mathbf{u}} & \sum_{j \in \mathcal{V}}\left\|x_j(T)-x_{j, \text { depart }}\right\|_2^2 \\
\text { s.t. } \quad & u_j(t)=0, t\notin\left[t_{j, \text { arrival }}, t_{j, \text { depart }}\right), j \in \mathcal{V} \\
& \sum_{j\in\mathcal{V}}u_j(t)\leq P(t), t\in\mathcal{T}\\
& \underline{u}_j\leq u_j(t) \leq\Bar{u}_j, t\in\mathcal{T}, j\in\mathcal{V}\\
& x_j(0) = x_{j,initial}, j\in\mathcal{V}\\
& x_j(t)=x_j(t-1)+\delta u_j(t), t \geq 1
\end{align}
\end{subequations}
In this problem, the objective is to meet terminal energy demand of each EV. (2b) indicates that charging power is only available when the charging session is alive. $t_{j,arrival}$ and $t_{j,depart}$ represent the arrival time and expected departure time of the $j_{th}$ EV. (2c) and (2d) define total power constraint and charging rate constraint of individual EV, respectively. With (2e) and (2f), the battery state is initialized with $x_{j,initial}$, and updated with charging rate $u_j(t)$. $\delta$ denotes charging efficiency.

\begin{tcolorbox}[top=10pt, colback=white, colframe=black, colbacktitle=black, center, enhanced, breakable,
attach boxed title to top left={yshift=-0.1in,xshift=0.15in}, boxed title style={boxrule=0pt,colframe=white,}, title=LLM4EV]
{\bf System:} You are an AI assistant specialized in solving EV charging problems. You have been provided with a predefined function called solve\_EV() that is capable of addressing various EV charging problems:
\begin{lstlisting}
import numpy as np
    
def Solve_EV(num_of_vehicles, timesteps, initial_states, max_power, terminal_states, dept_time, power_capacity, plot_fig):
    x_terminal=cp.Parameter(num_of_vehicles, name='x_terminal')
    x0 = cp.Parameter(num_of_vehicles, name='x0')
    max_sum_u = cp.Parameter(name='max_sum_u')
    u_max = cp.Parameter(num_of_vehicles, name='u_max')
    x = cp.Variable((num_of_vehicles, timesteps+1), name='x')
    u = cp.Variable((num_of_vehicles, timesteps), name='u')

    x_terminal.value=terminal_states
    x0.value=initial_states
    max_sum_u.value = power_capacity
    u_max.value=max_power*np.ones((num_of_vehicles, ))

    obj = 0
    constr = [x[:,0] == x0, x[:,-1] <= x_terminal]

    for t in range(timesteps):
        constr += [x[:,t+1] == x[:,t] + u[:,t],
        u[:,t] <= u_max,
        u[:,t] >= 0,
        cp.sum(u[:,t]) <= max_sum_u,
        u[:,t] <= (t*np.ones_like(dept_time)<dept_time)*100.0+0.000001]
    obj -= cp.norm(x[:, -1]-x_terminal, 2)
    prob = cp.Problem(cp.Maximize(obj), constr)
    prob.solve()

    if plot_fig==True:
        plt.plot(x.value[0])
        plt.plot(u.value[0])
        plt.show()


    return x.value, u.value
\end{lstlisting}
When a user requests you to solve an EV charging problem, you should ask the user in natural language to provide the necessary parameters. Then, based on user's response, you should generate code to invoke solve\_EV() function. Here is an example: 

\begin{lstlisting}
num_of_vehicles = 3
timesteps = 10
initial_states = [0, 0, 0]
max_power = 10.0
terminal_states = [70, 50, 100]
dept_time = [8, 6, 10]
power_capacity = 20
solve_EV(num_of_vehicles, timesteps, initial_states, max_power, terminal_states, dept_time, power_capacity, plot_fig)
\end{lstlisting}

{\bf User:}\sethlcolor{green!20}\hl{Can you help me to schedule the charging of electric vehicles?}\\

{\bf Assistant:}\sethlcolor{blue!20}\hl{Of course, I'd be happy to help. To provide the best result, I need some specific details from you. Could you please tell me the following:\\
1. How many vehicles do you have that you need to charge?\\
2. Over how many timesteps are you planning to charge your vehicles? A timestep could be an hour, for example.\\
3. What are the initial states (charge levels) of your vehicles? Please list for each vehicle if more than one.\\
4. What's the maximum power each of your vehicles can consume?\\
5. What are the final desired states (charge levels) for each of your vehicles?\\
6. Until what timestep does each of your vehicles not need to start charging?
Please list for each vehicle if more than one.\\
7. What's the total power capacity you want to use for charging all your vehicles?\\
8. Finally, would you like a plot figure of the charging status?}\\

{\bf User:}\sethlcolor{green!20}\hl{1. five,\\
2. 20 hours,\\
3. They all start from zero,\\
4. 10,\\
5. 100,\\
6. [10, 12, 16, 18, 20],\\
7. 30.\\}

{\bf Assistant:} \sethlcolor{blue!20}\hl{Thank you for the information. Based on your input, we can utilize our EV charging solver as follows:}
\begin{lstlisting}
num_of_vehicles = 5
timesteps = 20
# starting from zero for all vehicles
initial_states = [0, 0, 0, 0, 0]
max_power = 10.0
# desired state is 100 for all vehicles
terminal_states = [100, 100, 100, 100, 100]
# staggered start times for charging
dept_time = [10, 12, 16, 18, 20] 
power_capacity = 30
Solve_EV(num_of_vehicles, timesteps, initial_states, max_power, terminal_states, dept_time, power_capacity)
\end{lstlisting}

\sethlcolor{gray!20}\hl{Python Interpreter Execution.}

{\bf Assistant:} \sethlcolor{blue!20}\hl{The optimal schedule is as follows: [[10, 10, 10, 10, 10, 10, 10, 10, 10, 10, 0, 0, 0, 0, 0, 0, 0, 0, 0, 0], [8.229, 8.229, 8.229, 8.229, 8.229, 8.229, 8.230, 8.230, 8.231, 8.232, 8.840, 8.853, 0, 0, 0, 0, 0, 0, 0, 0], [4.978, 4.978, 4.978, 4.978, 4.978, 4.978, 4.978, 4.978, 4.978, 4.978, 7.668, 7.667, 8.720, 8.720, 8.722, 8.726, 0, 0, 0, 0], [3.500, 3.500, 3.500, 3.500, 3.500, 3.500, 3.500, 3.500, 3.500, 3.500, 6.815, 6.814, 8.482, 8.482, 8.483, 8.484, 8.719, 8.722, 0, 0], [2.347, 2.347, 2.347, 2.347, 2.347, 2.347, 2.347, 2.347, 2.347, 2.346, 5.093, 5.092, 8.082, 8.082, 8.082, 8.083, 8.459, 8.461, 8.546, 8.540]].}

\end{tcolorbox}

\subsection{Energy Context-based Document Comprehension}

\begin{figure*}
\centering
    \begin{AIbox}{LLM4Doc}
    {\small {\bf Question:} \sethlcolor{green!20}\hl{Please summarize this file.}}\\
    \parbox[t]{0.3\linewidth}{
    {\small {\bf Directly ask questions.}}\\
    
    {\small\bf Answer:} \scriptsize \begin{alltt}
    This document is an appendix to the NERC Rules of Procedure for Dispute Resolution. It contains the procedures for filing a complaint with the NERC Dispute Resolution Panel, including the process for filing a complaint, the timeline for the complaint process, and the requirements for filing a complaint. It also outlines the process for responding to a complaint, the timeline for the response process, and the requirements for responding to a complaint. It also provides a guideline for the 2023 DER\_A Model, outlining the requirements for the model, including the necessary components, safety protocols, and testing procedures.
    \end{alltt}
    \vspace{0.2in}
    {\small {\bf RAG}}\\
    
    {\small\bf Answer:} \scriptsize \begin{alltt}
    This document provides information from the Federal Energy Regulatory Commission (FERC) Home Page on the Internet, which is available on eLibrary. The full text of the document is available in PDF and Microsoft Word format for viewing, printing, and/or downloading. To access the document, the user must type the docket number excluding the last three digits in the docket number field. The document directs NERC to consider in its standards development process whether to include in new or modified Reliability Standards a requirement that planning coordinators and transmission planners include a wide set of grid stress performance conditions (i.e., both typical and extreme conditions) in planning assessments, and whether to require planning coordinators and transmission planners to account in planning assessments for both on-peak and off-peak conditions, normal and extreme. An example of a transmission system plan and performance requirements for extreme weather is provided, with Order No. 896, 88 FR 41262 (June 23, 2023), 183 FERC u00b6 61,191, at P 198 (2023).
    \end{alltt}}
    \hspace{0.2in}
    \parbox[t]{0.3\linewidth}{
    {\small {\bf RAG + prompt engineering}}\\
    
    {\small\bf Answer:} \scriptsize \begin{alltt}
    The document addresses concerns raised by commenters regarding data accessibility and burden associated with providing unregistered IBR (\textcolor{red}{In-Balance Resettlement}) modeling data and parameters. It acknowledges that transmission owners may face limitations in obtaining the necessary data for the reliable operation of the Bulk-Power System. Similar challenges may also be encountered by distribution providers in gathering adequate modeling data and parameters from IBR-DERs (\textcolor{red}{In-Balance Resettlement} Distributed Energy Resources). To mitigate these issues, the document suggests that existing processes of transmission planners, such as the New York Independent System Operator (NYISO) and CAISO, could be utilized to obtain specific information as required. It emphasizes the importance of demonstrating the need for information and ensuring proper protection and non-disclosure agreements are in place. The document also discusses the effective date and congressional notification of the rule, specifying that it will go into effect 60 days after publication in the Federal Register. It clarifies that the rule does not meet the definition of a “major rule” as per section 351 of the Small, implying the involvement of the Office of Information and Regulatory Affairs of the OMB (Office of Management and Budget) in assessing this determination. In summary, the document addresses concerns regarding data access and burden related to unregistered IBR modeling data and parameters. It proposes leveraging existing processes of transmission planners, highlights challenges faced by transmission owners and distribution providers, and provides information on the rule’s effective date and regulatory classification.
    \end{alltt}
    }
    \hspace{0.2in}
    \parbox[t]{0.3\linewidth}{
    {\small {\bf Human writing}}\\
    
    {\small\bf Answer:} \scriptsize \begin{alltt}
    FERC Order No. 901 addresses specific matters pertaining to the impacts of inverter-based resources (IBR) on the reliable operation of the Bulk-Power System. With an unprecedented proportion of nonsynchronous generation resources which employ inverters, rectifiers, and converters projected to connect over the next decade, this order is aims at providing system planners and operators the necessary tools to plan for and reliably integrate IBRs into the Bulk-Power System or to plan for IBRs connected to the distribution system that in the aggregate have a material impact on the Bulk-Power System. This order directs NERC to develop new or modified Reliability Standards to address reliability gaps pertaining to IBRs in four areas: data sharing; model validation; planning and operational studies; and performance requirements. Such Standards should ensure that NERC registered entities have the necessary data to predict the behavior of all IBRs. They should also ensure that IBR models are comprehensive, validated, and updated in a timely manner. In addition, the standards ensure that validated IBR models are included in transmission planning and operational studies to assess the reliability impacts on Bulk-Power System performance. And the standards should require that registered IBRs provide frequency and voltage support during frequency and voltage excursions. The order also provides guidelines on Informational Filing and Reliability Standard Development Timeline.
    \end{alltt}}

    \end{AIbox}
% \caption{Answers on GQ 1, generated under different experimental setup.}
\label{fig:LLM4Doc:GQ1}
\end{figure*}

\textbf{General questions} care more on the entire content of the document:

\begin{enumerate}
    \item Please summarize this file.
    \item What is the structure of this file?
    \item Please give me several key words strongly related to the content in this file.
\end{enumerate}

\noindent\textbf{Technical questions} care more on the technical details mentioned in the document:

\begin{enumerate}
    \item What is the Synchronization in this file?
    \item What is the phase-lock loop Synchronization in this file?
    \item Please summarize the main content of Synchronization part in this file.
    \item Please give an example in real-world application of the synchronization mentioned in this file.
    \item Please tell me, in Reliability Standards in this file, which standard is the most useful one?
\end{enumerate}

A comparison of summaries provided by the direct answering scheme, RAG scheme, RAG plus prompt engineering, and expert human writters are provided in the following text block using GPT-4. We note that answers provided by the LLMs are not always accurate, and the incorrect parts are market in red texts.

\end{document}